\documentclass[usenatbib]{mnras}
\usepackage{rotating}
\usepackage{multirow}
\usepackage{amssymb}
\usepackage{amsmath}
\usepackage{hyperref}
\usepackage{color}
\usepackage{placeins}

\def\apj{ApJ}
\def\apjl{ApJ}
\def\mnras{MNRAS}
\def\pasp{PASP}

\def\aap{A\&A}
\def\aj{AJ}
\def\apjs{APJS}

\def\araa{ARAA}

\title[Galaxy cluster luminosities and colours]{Galaxy cluster luminosities and colours, and their dependence on cluster mass and merger state}

\author[S. L. Mulroy et al.]
{Sarah L. Mulroy,$^{1}$\thanks{E-mail: smulroy@star.sr.bham.ac.uk} Sean L. McGee,$^{1}$ Steven Gillman,$^{1}$ Graham P. Smith,$^{1}$
\and
Chris P. Haines,$^{2}$ Jessica D\'emocl\`es,$^{1,3}$ Nobuhiro Okabe,$^{4,5}$ and Eiichi Egami$^{6}$
\vspace{2mm}\\
	$^1$ School of Physics and Astronomy, University of Birmingham, Edgbaston, Birmingham, B15 2TT, UK \\
    $^2$ INAF - Osservatorio Astronomico di Brera, Via Brera 28, 20122 Milano, Italy \\
	$^3$ Service d'Astrophysique AIM, CEA-Saclay, F-91191 Gif sur Yvette \\
    $^4$ Department of Physical Science, Hiroshima University, 1-3-1 Kagamiyama, Higashi-Hiroshima, Hiroshima 739-8526, Japan \\
    $^5$ Hiroshima Astrophysical Science Center, Hiroshima University, Higashi-Hiroshima, Kagamiyama 1-3-1, 739-8526, Japan \\
	$^6$ Steward Observatory, University of Arizona, 933 North Cherry Avenue, Tucson, AZ 85721, USA \\
}

\begin{document}

\date{Accepted. Received; in original form}

\pagerange{\pageref{firstpage}--\pageref{lastpage}} \pubyear{2017}

\maketitle

\label{firstpage}

\begin{abstract}
We study a sample of 19 galaxy clusters in the redshift range $0.15<z<0.30$ with highly complete spectroscopic membership catalogues (to $K < K^{\ast}(\rm z)+1.5$) from the Arizona Cluster Redshift Survey (ACReS); individual weak-lensing masses and near-infrared data from the Local Cluster Substructure Survey (LoCuSS); and optical photometry from the Sloan Digital Sky Survey (SDSS). We fit the scaling relations between total cluster luminosity in each of six bandpasses ({\it grizJK}) and cluster mass, finding cluster luminosity to be a promising mass proxy with low intrinsic scatter $\sigma_{\ln L|M}$ of only $\sim 10-20$ per cent for all relations. At fixed overdensity radius, the intercept increases with wavelength, consistent with an old stellar population. The scatter and slope are consistent across all wavelengths, suggesting that cluster colour is not a function of mass. Comparing colour with indicators of the level of disturbance in the cluster, we find a narrower variety in the cluster colours of `disturbed' clusters than of `undisturbed' clusters. This trend is more pronounced with indicators sensitive to the initial stages of a cluster merger, e.g. the Dressler Schectman statistic. We interpret this as possible evidence that the total cluster star formation rate is `standardised' in mergers, perhaps through a process such as a system-wide shock in the intracluster medium.
\end{abstract}


\section{Introduction}

The composition of galaxy clusters is thought to represent that of the whole Universe, and so they offer a window into astrophysics on both cluster and galaxy scales \citep[e.g.][]{Kravtsov2012}. Their position at the extreme end of the mass function makes them sensitive to the underlying cosmology and provides a late time estimate of the cosmological parameters, complementary to alternative probes such as the cosmic microwave background and supernovae \citep[e.g.][]{Weinberg2012}.

Accurate mass measurements of galaxy clusters are necessary to constrain the mass function, and thus cosmology \citep[e.g.][]{Allen2011}. Methods to make such measurements include: dynamical, which measure the depth of the potential well of the clusters using the velocities of the galaxies; hydrostatic, which assume that the gas pressure is balanced by the gravitational attraction; and gravitational weak-lensing, which measure the distortion of the light distribution from distant galaxies by the gravitational potential of the cluster.

While these methods each have different biases that require further exploration, well-constrained direct individual mass measurements require deep observations and extensive analysis that is not easily extended to very large samples. This motivates research into well calibrated scaling relations between easily measured `mass proxies' and cluster mass. The preferable scaling relation is one with minimal intrinsic scatter between observable and mass, and an observable that is easily obtainable from survey data.

Potential observables that could be suitable mass proxies cover a wide range of the electromagnetic spectrum, including: millimetre Sunyaev Zel'dovich effect \citep[e.g.][]{Arnaud2010,Marrone2012}, near-infrared luminosities \citep[e.g.][]{Lin2003,Mulroy2014}, optical measures such as richness \citep[e.g.][]{Rozo2009,Andreon2010} and velocity dispersion \citep[e.g.][]{Carlberg1997,Ruel2014}, and X-ray observables \citep[e.g.][]{Vikhlinin2006,Mantz2016}.

Promisingly, the work of \citet{Mulroy2014} showed total cluster near-infrared luminosity to be a low scatter mass proxy for a sample of clusters at $z\sim0.23$. Future wide field surveys will observe clusters at higher redshifts, where their rest frame optical light has been redshifted into the near-infrared filters. It is therefore important to determine whether the small scatter found in the near-infrared luminosity persists at bluer rest frame wavelengths.

Extending the study of total cluster luminosity to bluer bands also allows us to investigate the colour of a galaxy cluster. This colour corresponds to the average member galaxy colour, which in turn is an indicator of the age and metallicity of the stellar population within it. It has been known for some time that galaxies within galaxy clusters have old stellar populations, low current star formation rates (SFRs), and are relatively metal rich \citep{Nelan2005,vonderLinden2010,SmithR2012}. This highlights the influence of environment on galaxy properties, and motivates investigation into the galaxy populations within clusters of different evolutionary stages and morphological states.

The state of the cluster can be probed through central cluster properties. For instance, one common indicator of disturbance in the X-ray is cool core strength -- a measure of the rate of gas cooling in the centre of a cluster. A strong cool core suggests a more relaxed history \citep[e.g.][]{Poole2008,Rossetti2010}. Cluster mergers can disturb not only the cool cores but also the gravitational potential of a cluster, which can be seen in the dynamics of the cluster galaxies and probed through the bulk cluster properties \citep{DS1988,Burns1998}.

Here we combine weak-lensing mass measurements, which have been shown in simulations to be unbiased on average \citep{Oguri2011,Becker2011,Bahe2012}, with optical luminosities, which require only shallow imaging data. Optical luminosities have previously been shown to be good proxies for X-ray and dynamical mass measurements \citep[e.g.][]{Girardi2000,Popesso2005}. We utilise highly complete spectroscopic redshift catalogues in order to isolate issues arising from selecting members in colour-magnitude space, caused by the sensitivity of galaxy colour to astrophysics \citep[e.g.][]{Lu2009,Castignani2016}. We use the same member selection for every waveband to provide a clean probe of the underlying cluster physics.

In this paper we use a sample of 19 massive galaxy clusters to quantify the scaling relations between optical luminosities and weak-lensing mass, before investigating the trends between cluster colour and various indicators of the level of disturbance in these clusters. We introduce our data in Section \ref{sec:data}, present our results in Section \ref{sec:results} and our interpretation of cluster colour trends in Section \ref{sec:int}, before summarising in Section \ref{sec:summary}. All photometric measurements are in the AB system, and we assume $\Omega_{\rm M, 0}=0.3$, $\Omega_{\Lambda,0}=0.7$ and $H_0=70\,{\rm km\,s^{-1}\,Mpc^{-1}}$. In this cosmology, at the average cluster redshift, $\langle z \rangle=0.23$, 1 arcsec corresponds to a projected physical scale of 3.67 kpc.

\section{Data}\label{sec:data}

\subsection{Sample}\label{sec:sample}
The sample comprises 19 X-ray luminous galaxy clusters at $0.15<z<0.30$ (Table \ref{tab:sample}), which populate the overlap between three surveys:
the Sloan Digital Sky Survey (SDSS\footnote{\url{http://www.sdss.org/}}), a wide field photometric and spectroscopic survey;
the Local Cluster Substructure Survey (LoCuSS\footnote{\url{http://www.sr.bham.ac.uk/locuss}}) ``High-$L_X$'' sample, 50 well studied clusters from the multiwavelength survey of X-ray luminous clusters at $0.15<z<0.30$;
and the Arizona Cluster Redshift Survey (ACReS\footnote{\url{http://herschel.as.arizona.edu/acres/acres.html}}), a spectroscopic survey of 30 clusters drawn from the full LoCuSS sample. The LoCuSS ``High-$L_X$'' sample was selected on X-ray luminosity and the ACReS clusters are a representative sub-sample, while the overlap with SDSS is determined only by sky coverage. Thus, the main physical selection is on the X-ray luminosity.

\subsection{Cluster Luminosities}\label{sec:lums}
We have total $J$ and $K$ band Kron magnitudes for the cluster galaxies from LoCuSS, most from WFCAM on UKIRT, and two (ZwCl0857.9+2107 and Abell0963) from NEWFIRM on the Mayall 4-m telescope at Kitt Peak National Observatory \citep{Haines2009,Mulroy2014}.

All clusters in our sample also have SDSS Data Release 12 \textit{ugriz} band photometry \citep{Gunn1998,Doi2010,Alam2015}, from which we use the `modelmag' aperture magnitudes and `cmodelmag' total magnitudes. The $u$ band data with a magnitude limit of 22.0 is not deep enough to robustly measure the predominantly red cluster galaxies at these redshifts, so we discard this bluest band. All magnitudes are corrected for galactic extinction assuming the dust maps of \citet*{SFD1998}. 

To determine cluster membership we use spectroscopic information from MMT/Hectospec observations taken by ACReS \citep[][M.\ J.\ Pereira et al.\ in prep.]{Haines2013}. Cluster members are those galaxies within the characteristic cluster caustic in redshift-clustercentric radius space. The spectroscopic targeting was $K$ band limited (independent of colour) with a resulting average completeness of $\sim 75$ per cent for galaxies with $K < K^{\ast}(\rm z)+1.5$ within 1Mpc, and a weighting system was used to account for those objects not observed. This is calculated by weighting every potential spectroscopic target galaxy equally, then redistributing the weight from each galaxy lacking a redshift equally to its ten nearest neighbours on the sky that had the same priority level in the original targeting strategy.

\begin{figure}
  \centering
  \includegraphics[width=\linewidth]{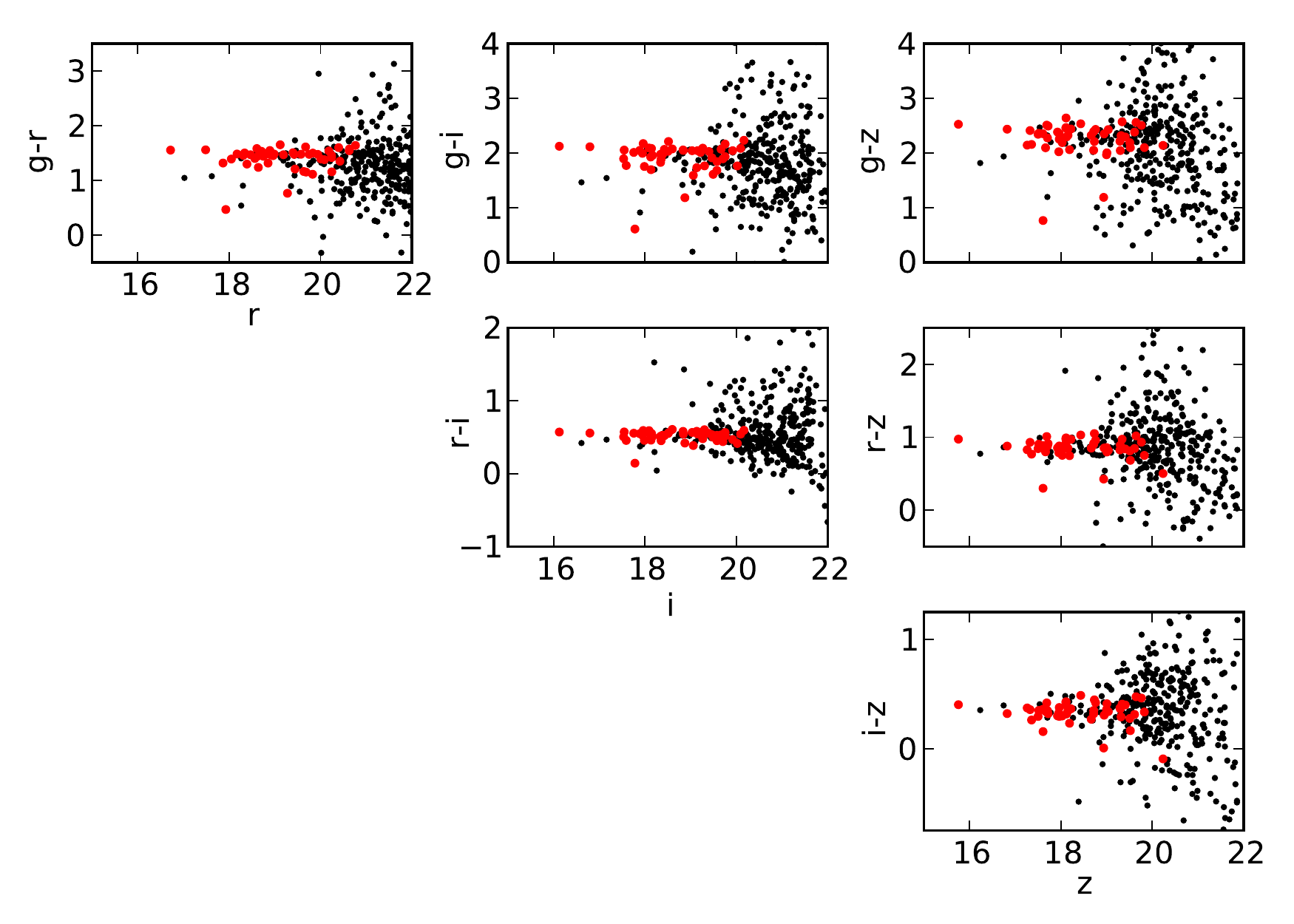}
  \caption{Colour-magnitude diagrams for our median mass cluster, Abell0068. Shown are all galaxies within 1Mpc of the cluster centre, with spectroscopically confirmed members marked in red.}
  \label{fig:colmag}
\end{figure}

In Figure \ref{fig:colmag}, we show SDSS colour-magnitude diagrams for a typical cluster in our sample (Abell0068). The spectroscopically confirmed cluster members are shown in red, and demonstrate the tight red sequence typical of massive galaxy clusters.

To convert the magnitudes to rest frame luminosities we apply $k$-corrections derived from the polynomial fitting functions of \citet{Chilingarian2010} and \citet{Chilingarian2012}, and normalise to solar luminosity \citep{Blanton2007}. At our redshifts, these fitting functions have been shown to agree on average with spectral energy distribution (SED) fitting programs \citep[e.g. \texttt{K-correct},][]{Blanton2007} to within $\sim0.02$ mags across the full range of optical and near-infrared data we use. We calculate the total cluster luminosity in each of the six bandpasses ($L_{\rm g},L_{\rm r},L_{\rm i},L_{\rm z},L_{\rm J},L_{\rm K}$) by summing the weighted luminosity of all cluster galaxies within a clustercentric radius derived from weak-lensing analysis (see Section \ref{sec:masses}) and with $K < K^{\ast}(\rm z)+1.5$, resulting in a roughly stellar mass limited selection of member galaxies. Uncertainties on the luminosities consist of two terms - one calculated by propagating the uncertainty on the weak-lensing radii, and the other from bootstrap resampling of the member galaxy luminosities.

\subsection{Cluster Masses}\label{sec:masses}
We use weak-lensing masses from \citet{Okabe2016}, where the authors used Subaru/Suprime-Cam imaging and fit an NFW \citep*{NFW1997} mass density profile to the weak shear profile of each cluster. $M_{\Delta}$ is the mass calculated within $r_{\Delta}$, the radius within which the average density is $\Delta \times \rho_{\rm crit}$, where $\rho_{\rm crit}=3H(z)^2/8\pi G$, the critical density of the Universe. We consider the overdensities $\Delta=\Delta_{\rm vir},500,2500$. $\Delta_{\rm vir}$ is defined as $\Delta_{\rm vir} = 18\pi^2 + 82x - 39x^2$ where $x = \Omega_{\rm M}(z) - 1$ \citep{Bryan&Norman1998}, and is equal to $\sim120$ at the average redshift of our sample. The weak-lensing error analysis is described fully in Section 3.1 of \citet{Okabe2016}, and includes shape noise, photometric redshift uncertainties, and uncorrelated large-scale structure.

\begin{table}
  \caption{Cluster sample}\label{tab:sample}
  \begin{center}
    \begin{tabular}{ l c c c }
    \hline
    \hline
    Cluster & RA & Dec & Redshift \\
    & $\alpha$ [J2000] & $\delta$ [J2000] & z \\
    \hline
    Abell0068 & 9.2785 & 9.1566 & 0.2546 \\
    ZwCl0104.4+0048 & 16.7057 & 1.0564 & 0.2545 \\
    Abell0267 & 28.1748 & 1.0072 & 0.2300 \\
    Abell0291 & 30.4296 & -2.1966 & 0.1960 \\
    Abell0586 & 113.0845 & 31.6335 & 0.1710 \\
    Abell0611 & 120.2367 & 36.0566 & 0.2880 \\
    Abell0697 & 130.7398 & 36.3666 & 0.2820 \\
    ZwCl0857.9+2107 & 135.1536 & 20.8946 & 0.2347 \\
    Abell0963 & 154.2652 & 39.0470 & 0.2060 \\
    Abell1689 & 197.8730 & -1.3410 & 0.1832 \\
    Abell1758N & 203.1600 & 50.5600 & 0.2792 \\
    Abell1763 & 203.8337 & 41.0012 & 0.2279 \\
    Abell1835 & 210.2588 & 2.8786 & 0.2528 \\
    Abell1914 & 216.4860 & 37.8165 & 0.1712 \\
    ZwCl1454.8+2233 & 224.3131 & 22.3428 & 0.2578 \\
    Abell2219 & 250.0827 & 46.7114 & 0.2281 \\
    RXJ1720.1+2638 & 260.0420 & 26.6260 & 0.1640 \\
    RXJ2129.6+0005 & 322.4165 & 0.0894 & 0.2350 \\
    Abell2390 & 328.4034 & 17.6955 & 0.2329 \\
    \hline
    \end{tabular}
  \end{center}
{\footnotesize}
\end{table}

\section{Results}\label{sec:results}

\subsection{Scaling Relations}\label{sec:scaling_relations}

We quantify the scaling relations between cluster luminosities $L$ and weak-lensing masses $M_{\rm WL}$ by performing linear regression on the logarithmic values using the method of \citet{Kelly2007}. The scaling relation is parameterised as:

\begin{equation}
  \frac{L}{10^{12}L_{\odot}}=a \left(\frac{M_{\rm WL}}{10^{15}M_{\odot}}\right)^{b},
\end{equation}
with intercept $a$, slope $b$, and intrinsic scatter $\sigma_{\ln L | M_{\rm WL}}$.

We do not consider selection effects in this work because the effects are diluted as a consequence of our sample being an overlap of several surveys. The LoCuSS ``High-$L_X$'' sample was selected on X-ray luminosity, and the overlap with ACReS and SDSS is not dependent on any cluster property. We note that the covariance between $L_{\rm{X,RASS}}$ and optical/near-infrared luminosity is expected to be minimal and lead to only minor selection effects (Mulroy et al., in prep.).

\begin{figure*}
  \centering
  \includegraphics[width=0.7\linewidth]{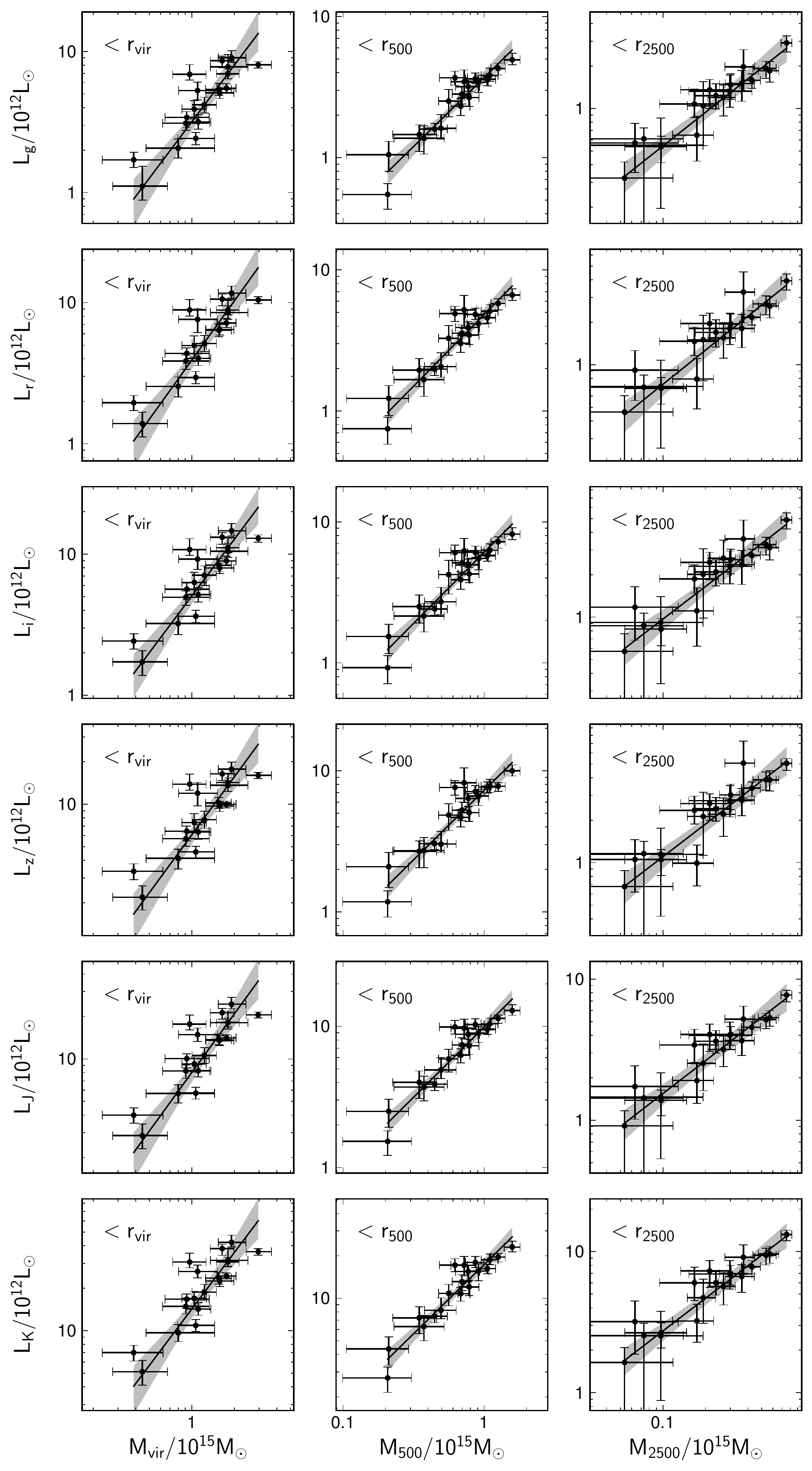}
  \caption{Scaling relations between the six total cluster luminosities and weak-lensing cluster mass, where we show the data points, resultant scaling relation and 68 per cent confidence region. Luminosities are calculated from a $K$ band limited sample of galaxies, and both luminosities and masses are measured within $r_{\rm vir}$ [left], $r_{500}$ [middle] and $r_{2500}$ [right].}
  \label{fig:relations}
\end{figure*}

\begin{figure}
  \centering
  \includegraphics[width=\linewidth]{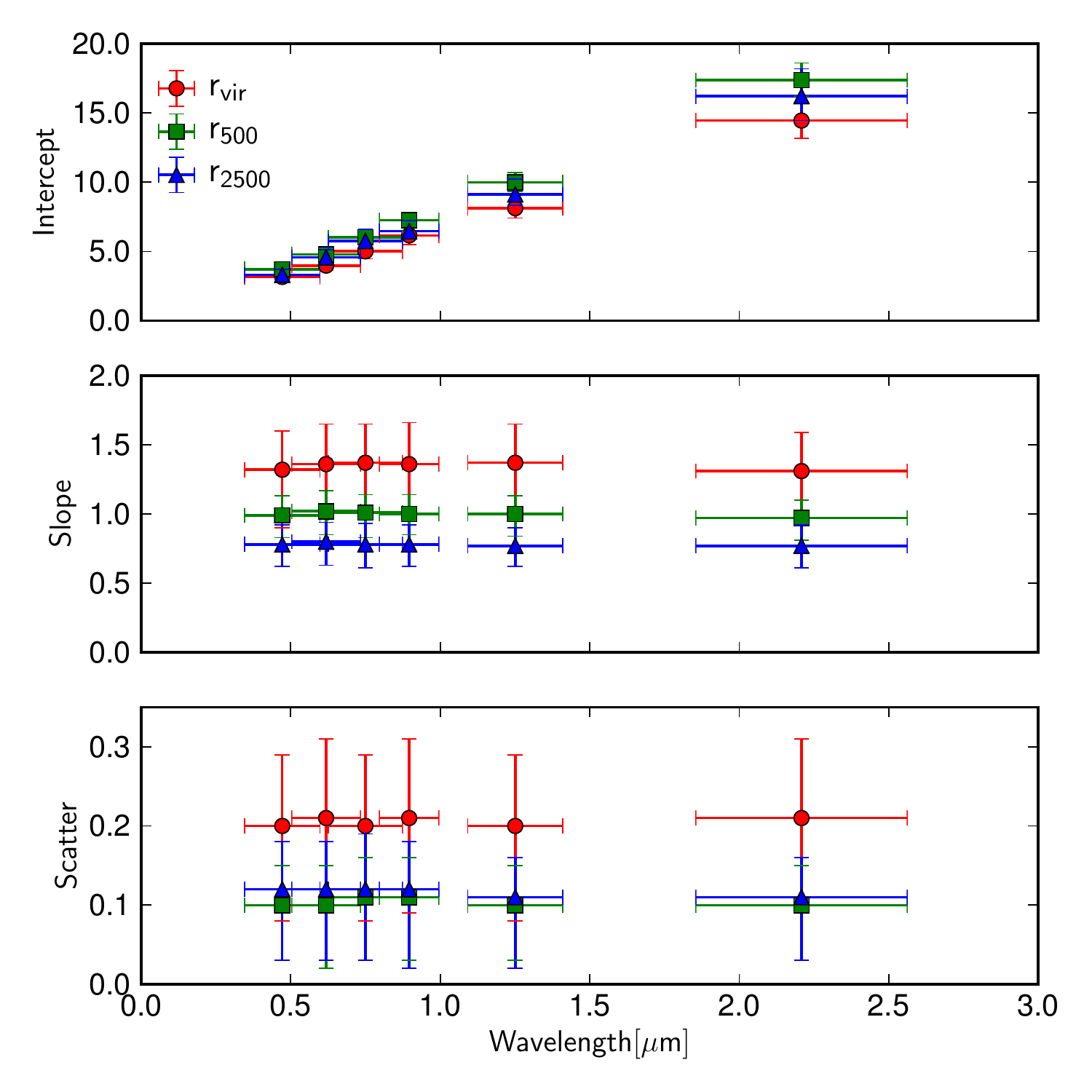}
  \caption{Scaling relation parameters (intercept [top], slope [middle] and intrinsic scatter [bottom]) as a function of the bandpass wavelength.}
  \label{fig:wavelengths}
\end{figure}

We perform linear regressions of the total cluster luminosities in 6 bandpasses (\textit{grizJK}) within 3 overdensity radii ($r_{\rm vir}$, $r_{500}$, $r_{2500}$) against the weak-lensing cluster masses within the same radii. In Figure \ref{fig:relations} we show the data points and resultant scaling relation (and 68 per cent confidence region) for each of these bandpass and radius combinations. The scaling relation parameters (intercept, slope, and intrinsic scatter) are shown in Table \ref{tab:fits}, and their trends with wavelength visually presented in Figure \ref{fig:wavelengths}. We note the following features in these results:

\begin{table}
  \caption{Scaling relation parameters}\label{tab:fits}
  \begin{center}
    \begin{tabular}{ l c c c }
\hline
\hline
Bandpass & Intercept & Slope & Scatter \\
 & $a$ & $b$ & $\sigma_{\ln L | M_{\rm WL}}$ \\
\hline
\multicolumn{4}{c}{$r_{\rm vir}$} \\
$L_g$  &  $ 3.16 ^ { + 0.39 } _ { - 0.34 } $  &  $ 1.32 ^ { + 0.28 } _ { - 0.42 } $  &  $ 0.20 ^ { + 0.09 } _ { - 0.12 } $ \\ 
$L_r$  &  $ 3.98 ^ { + 0.59 } _ { - 0.35 } $  &  $ 1.36 ^ { + 0.29 } _ { - 0.37 } $  &  $ 0.21 ^ { + 0.10 } _ { - 0.12 } $ \\ 
$L_i$  &  $ 5.01 ^ { + 0.61 } _ { - 0.55 } $  &  $ 1.37 ^ { + 0.28 } _ { - 0.37 } $  &  $ 0.20 ^ { + 0.09 } _ { - 0.12 } $ \\ 
$L_z$  &  $ 6.17 ^ { + 0.91 } _ { - 0.67 } $  &  $ 1.36 ^ { + 0.30 } _ { - 0.39 } $  &  $ 0.21 ^ { + 0.10 } _ { - 0.12 } $ \\ 
$L_J$  &  $ 8.13 ^ { + 1.20 } _ { - 0.72 } $  &  $ 1.37 ^ { + 0.28 } _ { - 0.39 } $  &  $ 0.20 ^ { + 0.09 } _ { - 0.12 } $ \\ 
$L_K$  &  $ 14.45 ^ { + 1.76 } _ { - 1.27 } $  &  $ 1.31 ^ { + 0.28 } _ { - 0.39 } $  &  $ 0.21 ^ { + 0.10 } _ { - 0.11 } $ \\ 
\hline
\multicolumn{4}{c}{$r_{500}$} \\
$L_g$  &  $ 3.72 ^ { + 0.27 } _ { - 0.25 } $  &  $ 0.99 ^ { + 0.14 } _ { - 0.16 } $  &  $ 0.10 ^ { + 0.05 } _ { - 0.07 } $ \\ 
$L_r$  &  $ 4.79 ^ { + 0.34 } _ { - 0.32 } $  &  $ 1.02 ^ { + 0.15 } _ { - 0.17 } $  &  $ 0.10 ^ { + 0.05 } _ { - 0.08 } $ \\ 
$L_i$  &  $ 6.03 ^ { + 0.43 } _ { - 0.04 } $  &  $ 1.01 ^ { + 0.13 } _ { - 0.18 } $  &  $ 0.11 ^ { + 0.05 } _ { - 0.08 } $ \\ 
$L_z$  &  $ 7.24 ^ { + 0.52 } _ { - 0.48 } $  &  $ 1.00 ^ { + 0.14 } _ { - 0.15 } $  &  $ 0.11 ^ { + 0.05 } _ { - 0.08 } $ \\ 
$L_J$  &  $ 10.00 ^ { + 0.72 } _ { - 0.67 } $  &  $ 1.00 ^ { + 0.13 } _ { - 0.16 } $  &  $ 0.10 ^ { + 0.05 } _ { - 0.07 } $ \\ 
$L_K$  &  $ 17.38 ^ { + 1.24 } _ { - 1.16 } $  &  $ 0.97 ^ { + 0.13 } _ { - 0.16 } $  &  $ 0.10 ^ { + 0.05 } _ { - 0.07 } $ \\ 
\hline
\multicolumn{4}{c}{$r_{2500}$} \\
$L_g$  &  $ 3.31 ^ { + 0.40 } _ { - 0.29 } $  &  $ 0.78 ^ { + 0.14 } _ { - 0.16 } $  &  $ 0.12 ^ { + 0.06 } _ { - 0.09 } $ \\ 
$L_r$  &  $ 4.57 ^ { + 0.56 } _ { - 0.50 } $  &  $ 0.80 ^ { + 0.14 } _ { - 0.17 } $  &  $ 0.12 ^ { + 0.06 } _ { - 0.09 } $ \\ 
$L_i$  &  $ 5.75 ^ { + 0.85 } _ { - 0.51 } $  &  $ 0.78 ^ { + 0.15 } _ { - 0.17 } $  &  $ 0.12 ^ { + 0.07 } _ { - 0.09 } $ \\ 
$L_z$  &  $ 6.46 ^ { + 0.79 } _ { - 0.57 } $  &  $ 0.78 ^ { + 0.14 } _ { - 0.16 } $  &  $ 0.12 ^ { + 0.06 } _ { - 0.10 } $ \\ 
$L_J$  &  $ 9.12 ^ { + 1.11 } _ { - 0.80 } $  &  $ 0.77 ^ { + 0.13 } _ { - 0.15 } $  &  $ 0.11 ^ { + 0.05 } _ { - 0.09 } $ \\ 
$L_K$  &  $ 16.22 ^ { + 1.98 } _ { - 1.76 } $  &  $ 0.77 ^ { + 0.15 } _ { - 0.16 } $  &  $ 0.11 ^ { + 0.05 } _ { - 0.08 } $ \\ 
\hline
    \end{tabular}
  \end{center}
{\footnotesize}
\end{table}

\FloatBarrier

\begin{figure}
  \centering
  \includegraphics[width=\linewidth]{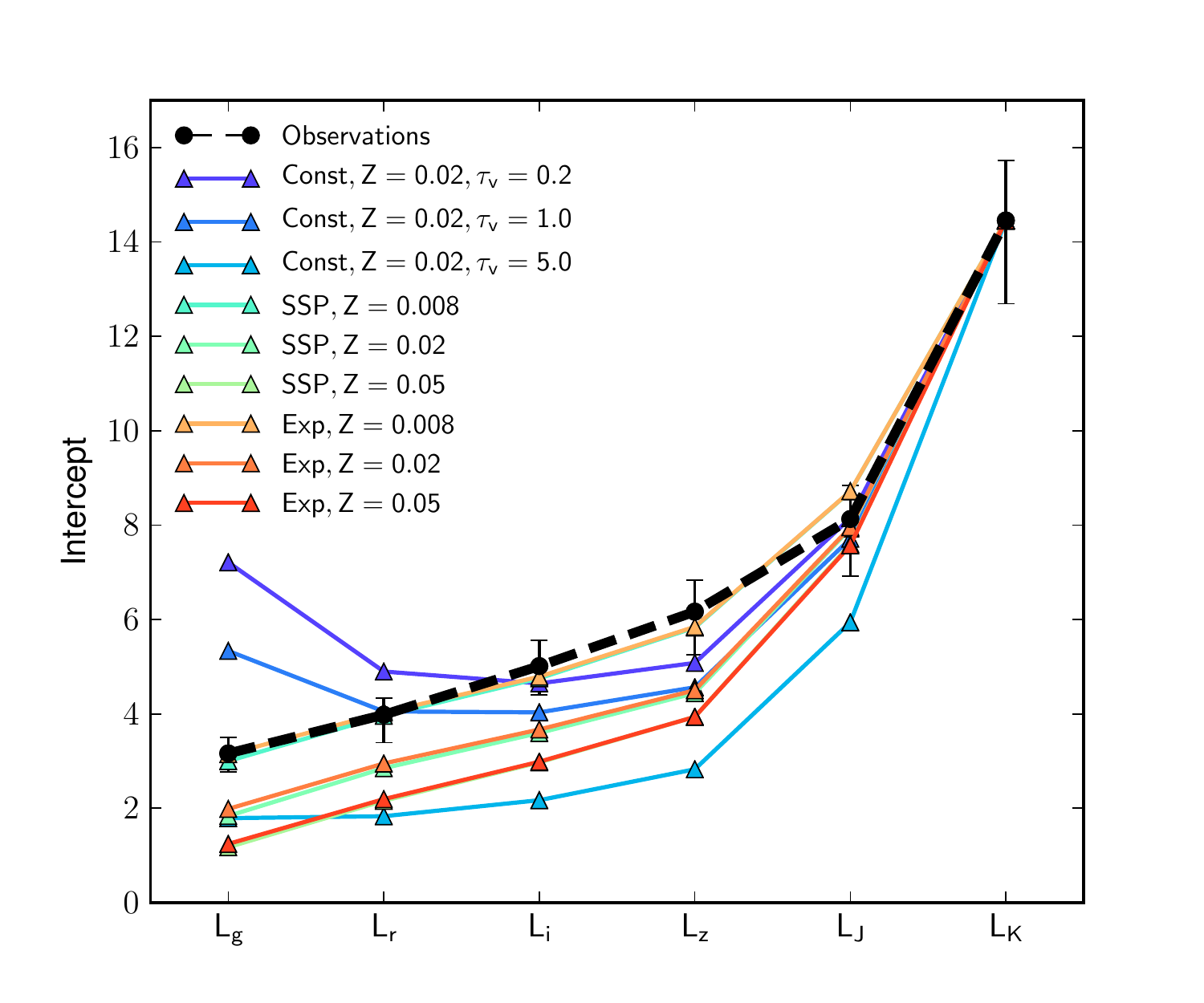}
  \caption{Comparison of observed intercept (black dashed line) with predicted intercept values from updated \citet{BC2003} stellar population models (blue lines: constant star formation history model, green lines: single stellar population model, red lines: exponentially decaying model), showing a trend of increasing intercept with increasing wavelength for all models and the observations. All models normalised with respect to the $K$ band value of the observations.}
  \label{fig:intercept}
\end{figure}

\begin{enumerate}
\item At fixed radius, both the slope and scatter of the scaling relations are consistent across the wavelength range ($\sim 0.47 - 2.21 \mu \rm{m}$). The same trend is found by \citet{Popesso2005}, although our absolute values do not always agree. Some difference is expected due to a different mass measurement method and linear regression scheme. We revisit this observed trend in Section \ref{sec:colour}.

\item At fixed radius, the intercept increases with increasing wavelength, as shown in the top panel of Figure \ref{fig:wavelengths}. This rising intercept is a reflection of the SED of the cluster population, which predominantly consists of red galaxies. To understand the shape further we present Figure \ref{fig:intercept}, which shows the intercept values at each bandpass for the relations measured within $r_{\rm vir}$, and compare these to values assuming updated \citet{BC2003} stellar population models calculated using the tool \texttt{EzGal} \citep{EzGal} and normalised to the $K$ band of the observations.

A constant star formation history model (blue lines) for solar metallicity ($Z=0.02$) and the common assumption of dust extinction \citep[$\tau_v$ = 1,][]{Brinchmann2004, Garn2010} is a reasonable match to the observed data in redder bands, but diverges in the bluer bands. Decreasing the dust extinction in such a model ($\tau_v$ = 0.2) improves agreement at the red end but increases the discrepancy at the blue end, while increasing the dust extinction ($\tau_v$ = 5) improves agreement with the bluest band but is consistently below the observed data. It is not possible to fit the observed data with this model by varying the dust extinction due to the shape of the predictions from a constant star formation history model.

We find that a single stellar population model (SSP, green lines) and an exponentially decaying model (with a timescale of 1 Gyr, red lines) are almost indistinguishable, and similar in shape to the observed data. These models with solar metallicity agree well with the observed data at the red end, while better agreement across all bands can be found if we allow the metallicity of the galaxies to vary. The observational intercepts of the relations can be well reproduced with either a single stellar population or quickly decaying exponential, with metallicity of $\sim0.4$ solar. This is in good agreement with previous studies of the stellar populations of low redshift cluster galaxies \citep{Nelan2005, Pasquali2010, SmithR2012}.

\item For all relations, the scatter is higher within $r_{\rm vir}$ than within $r_{500}$ or $r_{2500}$. This is consistent with the increased volume at larger radii allowing for more variation in the large scale structure found within the outskirts of the cluster. As a simple example, infalling groups are more likely to be found in the region between $r_{500}$ and $r_{\rm vir}$, and with greater variation than within $r_{500}$. Looking at galaxy clusters in the Millennium N-body dark matter simulation \citep{Springel2005, McGee2009}, we found that at fixed cluster mass, the fractional scatter in the number of galaxies within $r_{\rm vir}$ is roughly double that within both $r_{500}$ and $r_{2500}$, consistent with our observations.

A related effect is the fraction of `interlopers' -- spectroscopically confirmed members that are projected onto the cluster but lie beyond the physical radius of the cluster. Given that interlopers are largely uncorrelated with the cluster, a larger interloper fraction suggests a larger variation in that fraction, and therefore a larger inferred scatter. The fraction of members that have not yet passed within $r_{200}$, quantified using the Millennium simulation as in \citet{Haines2013}, is 3.44, 8.76 and 21.14 per cent within projected $r_{2500}$, $r_{500}$ and $r_{\rm vir}$ respectively, consistent with our observed larger scatter within $r_{\rm vir}$. These interlopers are likely to be bluer than a typical cluster galaxy, and so will slightly increase the intercepts of our relations particularly in the blue bands. This effect will be small, and we don't expect it to affect the results of our analysis in Section \ref{sec:colour} as we work within a single overdensity radius.

\item At fixed wavelength, the slope increases with increasing radius. This trend is much less prominent when we repeat the analysis without the brightest cluster galaxy (BCG) luminosity, with slopes increasing by $\sim0.05$, $\sim0.1$ and $\sim0.2$ within $r_{\rm vir}$, $r_{500}$ and $r_{2500}$ respectively, resulting in broad agreement in the derived slopes. This suggests that the slope for centrals is shallower and more of a dominant factor at smaller radii. Indeed, this is in agreement with theoretical models that find the stellar mass of a BCG in a cluster of this size does not strongly scale with cluster mass \citep{Behroozi2013, McCarthy2017}. At larger radii, the luminosity of satellite galaxies (which scales strongly with cluster mass) makes up a larger fraction of the total luminosity and thus drives the slope to be steeper. The remaining trend could be explained if the mass-concentration relation of satellite galaxies is shallower than that of the dark matter.

\item The low intrinsic scatter across all wavelengths means that optical/near-infrared light is a good mass proxy for upcoming surveys, as discussed in Section \ref{sec:summary}.
\end{enumerate}

\subsection{Cluster Colour}\label{sec:colour}

\begin{figure*}
  \centering
  \includegraphics[width=\linewidth]{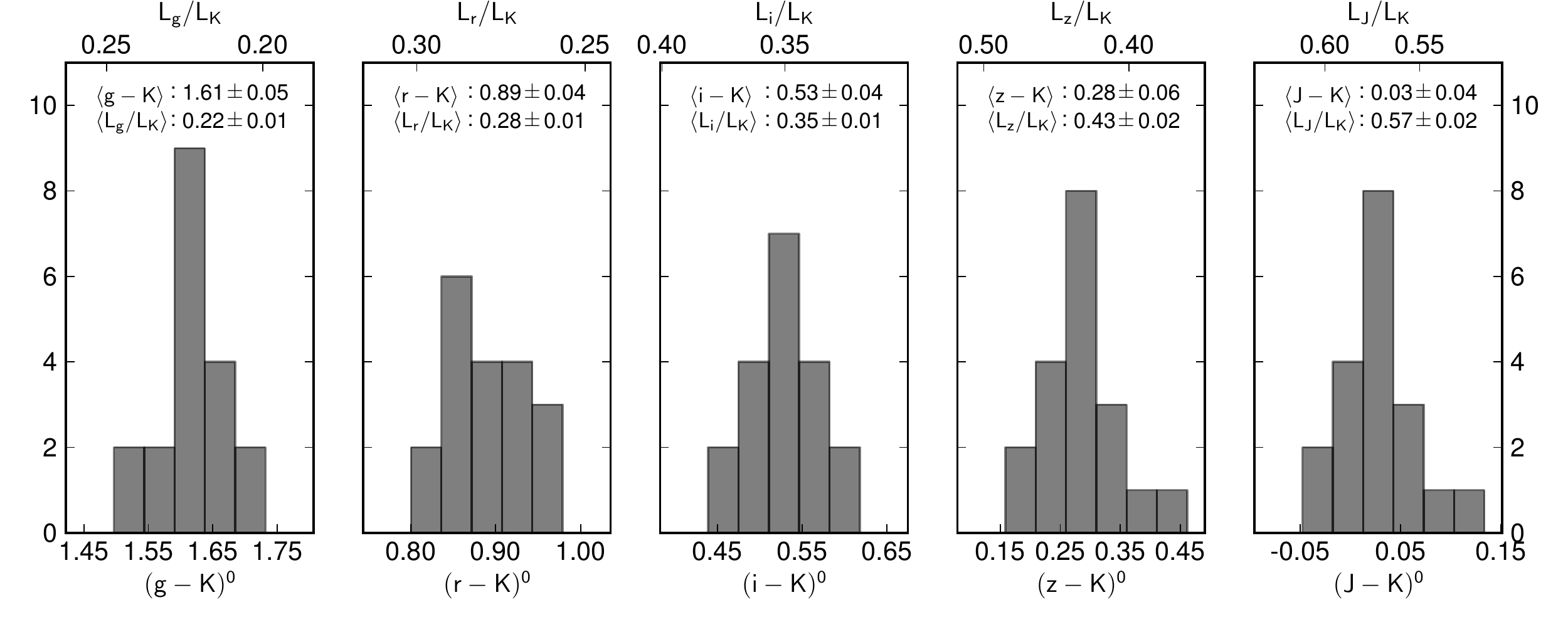}
  \vspace*{-6mm}
  \caption{Histogram of the rest frame cluster colours within $r_{\rm vir}$.}
  \label{fig:hist}
\end{figure*}

In Section \ref{sec:scaling_relations}, we found that the relation between total luminosity and cluster mass has the same slope across all wavelengths (within a given radius), which suggests that the colour of clusters is not a function of mass. Further, the low scatter in the relations places an upper limit on the variability of the colour of clusters.
As shown in Figure \ref{fig:hist}, we find variability in the cluster colours (standard deviation of the distribution of cluster colours) within $r_{\rm vir}$  on the scale of $\sigma \sim 0.05$ magnitudes in the full range of colours.

To understand this level of variation, we use updated \citet{BC2003} models with a single stellar population, which gave a reasonable match to the intercept of the scaling relations. For an SSP model with a fixed age of 10 Gyr, sampling the galaxy metallicity uniformly in the logarithm between 0.4 solar and 1 solar, leads to an average ($g-K$) colour variation of $\sigma \sim 0.05$ magnitudes (and similar in other bands). Similarly, at fixed solar metallicity, sampling SSP age between 7 Gyr and 10 Gyr leads to an average colour variation of $\sigma \sim 0.05$ magnitudes. It is worth noting that while this colour variation and the required change in metallicity or age is moderate in terms of individual galaxies, we are considering the mean for each cluster population, for which it is a large variation and requires further investigation to understand.

To investigate the source of this variability we explore the correlation between cluster colour and various indicators of the level of disturbance in that cluster. We consider seven indicators, four of which trace the bulk cluster properties and three of which are driven by the properties of the central region of the cluster.

\subsubsection{Bulk Cluster Properties}\label{sec:bulk}

The DS statistic \citep{DS1988} is a substructure test similar to a $\chi^2$ statistic that quantifies local deviations in mean velocity $\overline{\nu}$ and velocity dispersion $\sigma$. For each cluster member the local $\overline{\nu}$ and $\sigma$ are calculated using $N_{\rm{nn}}$ nearest neighbours, and compared to the global cluster values:
\begin{equation}
\delta_i^{2} = (N_{\rm{nn}}/\sigma^{2})\left[\left(\overline{\nu} - \overline{\nu}_{i, \text{local}}\right)^{2} + \left(\sigma - \sigma_{i, \text{local}}\right)^{2}\right].
\label{ds}
\end{equation}

\noindent The DS statistic, $\Delta_{\rm DS}$, is the sum of $\delta$, and after being normalised by the number of cluster galaxies is $\sim 1$ for clusters with a Gaussian velocity distribution, with higher values indicating the presence of substructure.

We calculate this statistic considering all members within $r_{\rm vir}$, and using $N_{\rm nn} = \sqrt{N_{\rm members}}$ to keep the measurement consistent between clusters of varying richness, although our measurements are not significantly affected by choosing a fixed number within this range. To quantify the statistical significance of this measurement we also calculate the P-value, by repeating the measurement after random reassignments of member positions to velocities. The P-value is the fraction of times this reassigned measurement is greater than the original DS statistic. We find only three clusters with a P-value $> 0.01$, corresponding to the three smallest DS statistics, and exclude these values from our analysis.

We define the magnitude gap, $\Delta M_{1,2}$, as the difference in $K$ band magnitude between the two $K$ band brightest cluster members within $0.5 \rm{r_{vir}}$. This gives an indication of the time since the last major merger activity in a cluster; a smaller gap suggests more recent infall of bright galaxies, while a larger gap suggests that the bright central galaxies have had time since any significant merger event to accrete onto the BCG \citep[e.g.][]{Dariush2010,Deason2013}.

We also calculate the projected separation between the X-ray centroid \citep{Martino2014} and the BCG position, $\Delta^{\rm{BCG}}_{\rm{X\mbox{-}ray}}$. In a dynamically relaxed cluster both the X-ray emitting hot gas and the BCG are centred on the minimum of the gravitational potential well, and so a larger separation indicates a more disturbed cluster.

Finally, we use the centroid shift parameter, $\langle w \rangle$, calculated in \citet{Martino2014} as a measure of the cluster X-ray morphology. It is defined as the standard deviation of the projected separation between the X-ray peak and the X-ray centroid calculated in circular apertures in the range $[0.05-1]r_{500}$. Clusters with high centroid shift are typically disturbed clusters, while those with low centroid shift are typically more relaxed. We note that as both $\Delta^{\rm{BCG}}_{\rm{X\mbox{-}ray}}$ and $\langle w \rangle$ are projected separations, they are insensitive to separation along the line of sight.

\subsubsection{Central Cluster Properties}\label{sec:central}

In the centre of some clusters, the intracluster medium (ICM) is strongly radiating and cooling. The cores of these clusters are therefore cool and dense, with low entropy and high surface brightness \citep[e.g.][]{Poole2008,Rossetti2010}. We use three parameters to probe the presence of these cool cores, and therefore to indicate the dynamical state of the ICM.

Following \citet{Santos2008} we calculate the surface brightness concentration, $c_{SB}$, as the ratio of the peak central surface brightness and the ambient surface brightness:

\begin{equation}
    c_{SB} = \frac{\rm{SB}(<40\rm{kpc})}{\rm{SB}(<400\rm{kpc})}.
\end{equation}

\noindent The surface brightness probes the emission of the ICM, and so a higher surface brightness concentration suggests the presence of a cool core, and therefore a more relaxed system.

\citet{Sanderson2009} calculate $\alpha$, the logarithmic slope of the gas density profile at $0.04r_{500}$ ($\sim 40\rm{kpc}$ for these objects) for all but one (Abell0291) of the clusters in our sample. The gas density slope traces the temperature slope, which steepens with increased cooling, and so a more negative $\alpha$ implies stronger cooling.

Also from \citet{Sanderson2009}, we use central entropy, $K$, measured within $20\rm{kpc}$ and defined as $K=Tn_{e}^{-2/3}$, where T is the cluster temperature and $n_{e}$ is the electron density. As a measure of the thermal history of the ICM, lower entropy is associated with the presence of a cool core.

\subsubsection{Cluster Colour Trends}\label{sec:colour_trends}

\begin{figure}
  \centering
  \includegraphics[width=0.85\linewidth]{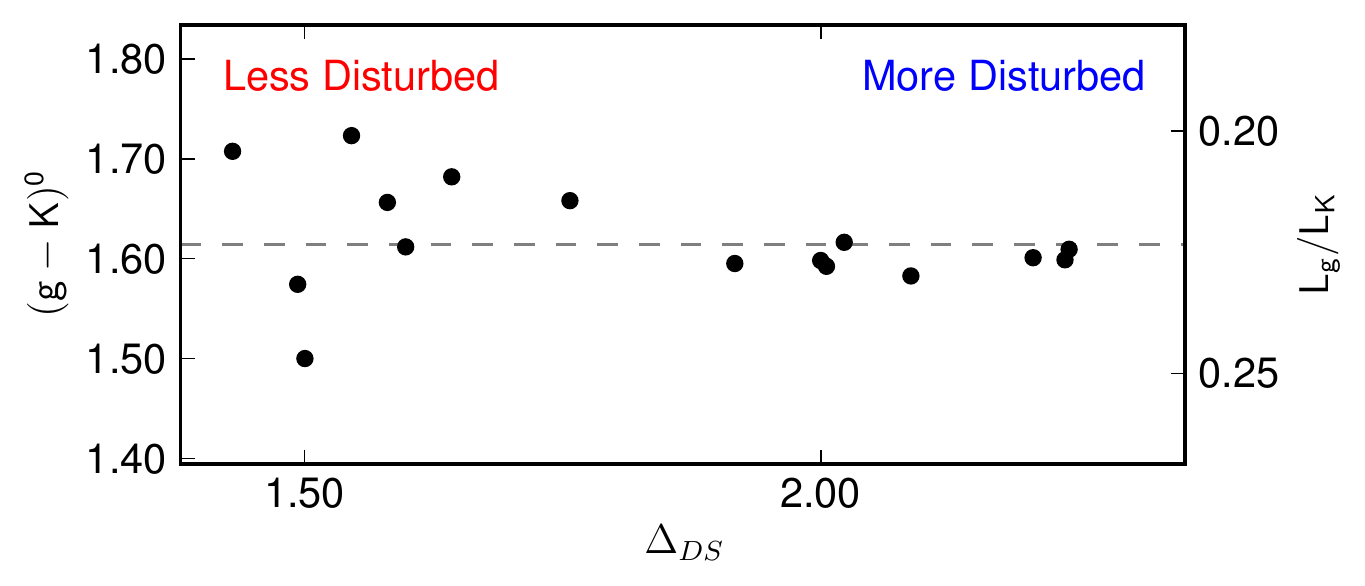}
  \includegraphics[width=0.85\linewidth]{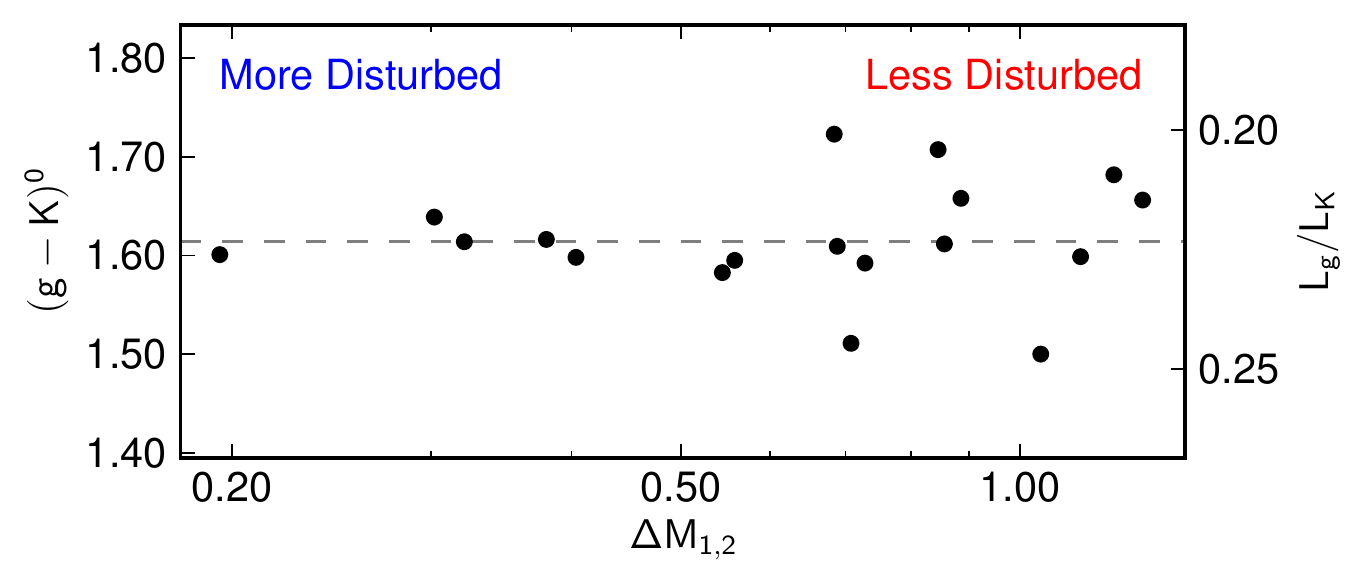}
  \includegraphics[width=0.85\linewidth]{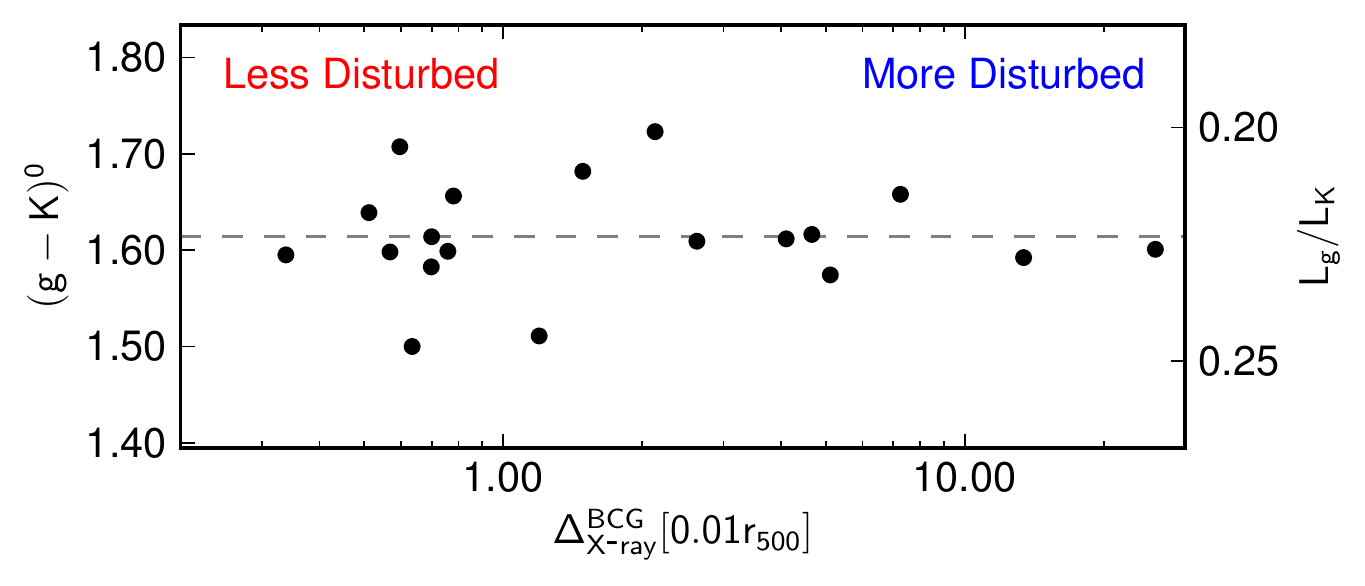}
  \includegraphics[width=0.85\linewidth]{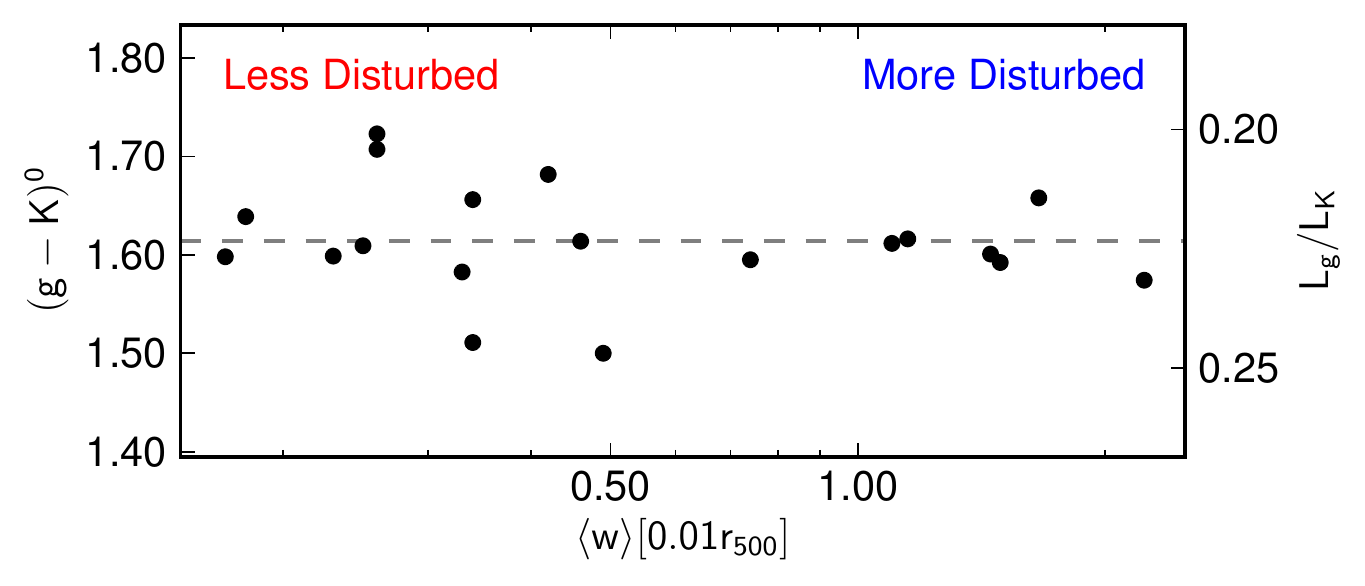}
  \includegraphics[width=0.85\linewidth]{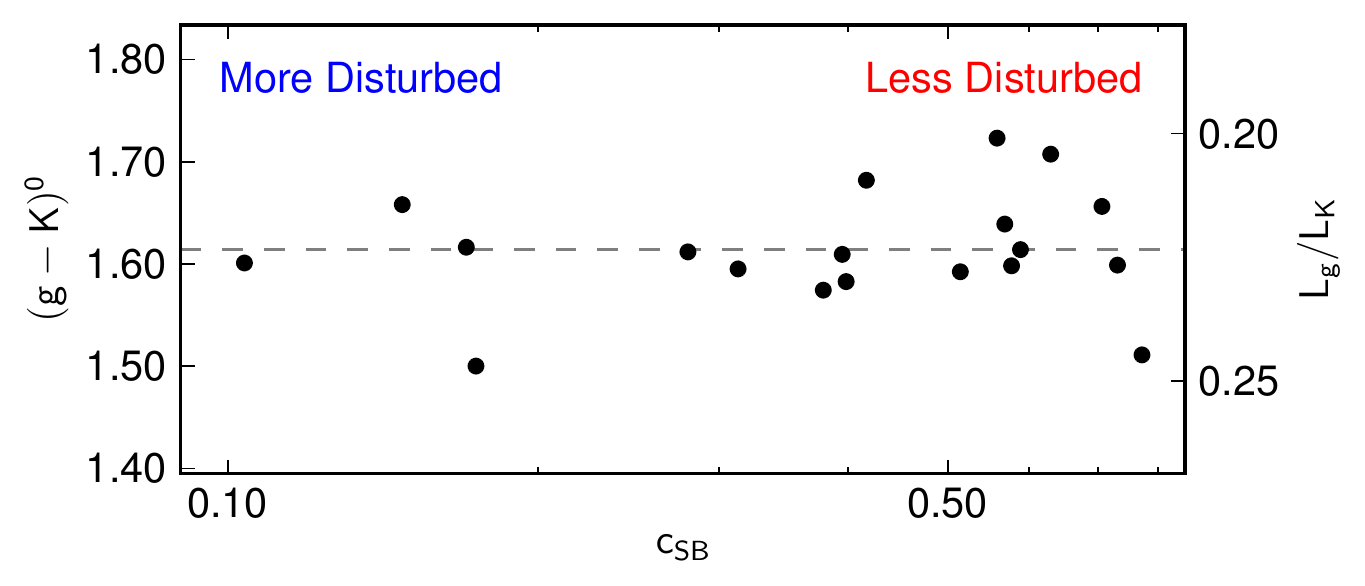}
  \includegraphics[width=0.85\linewidth]{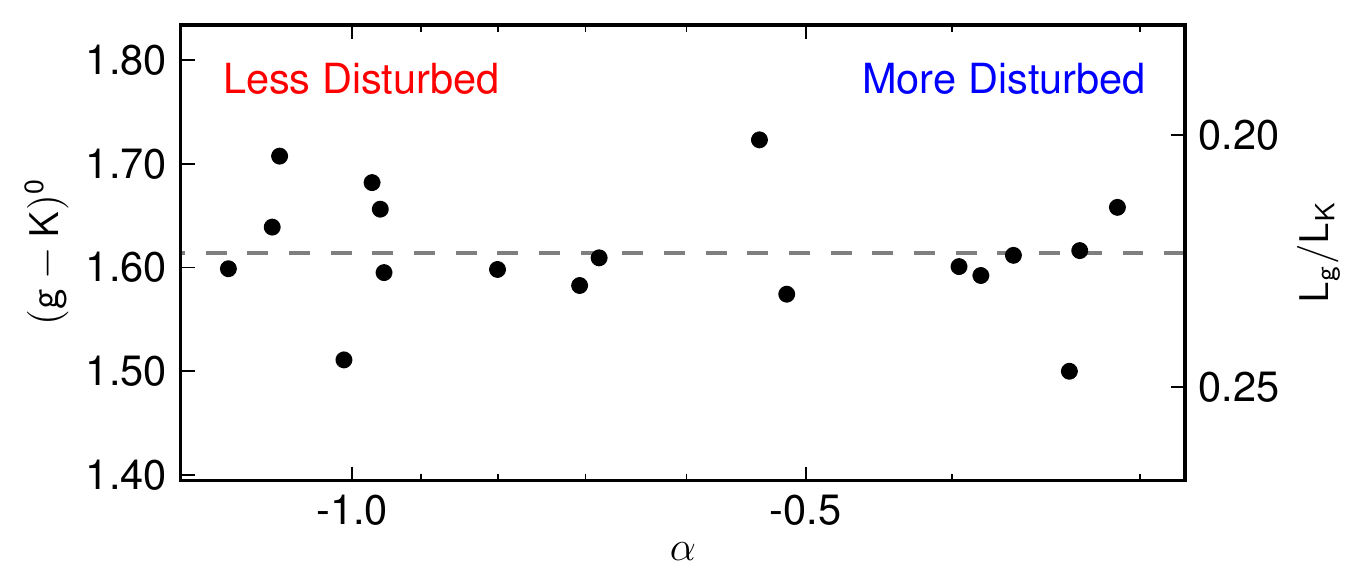}
  \includegraphics[width=0.85\linewidth]{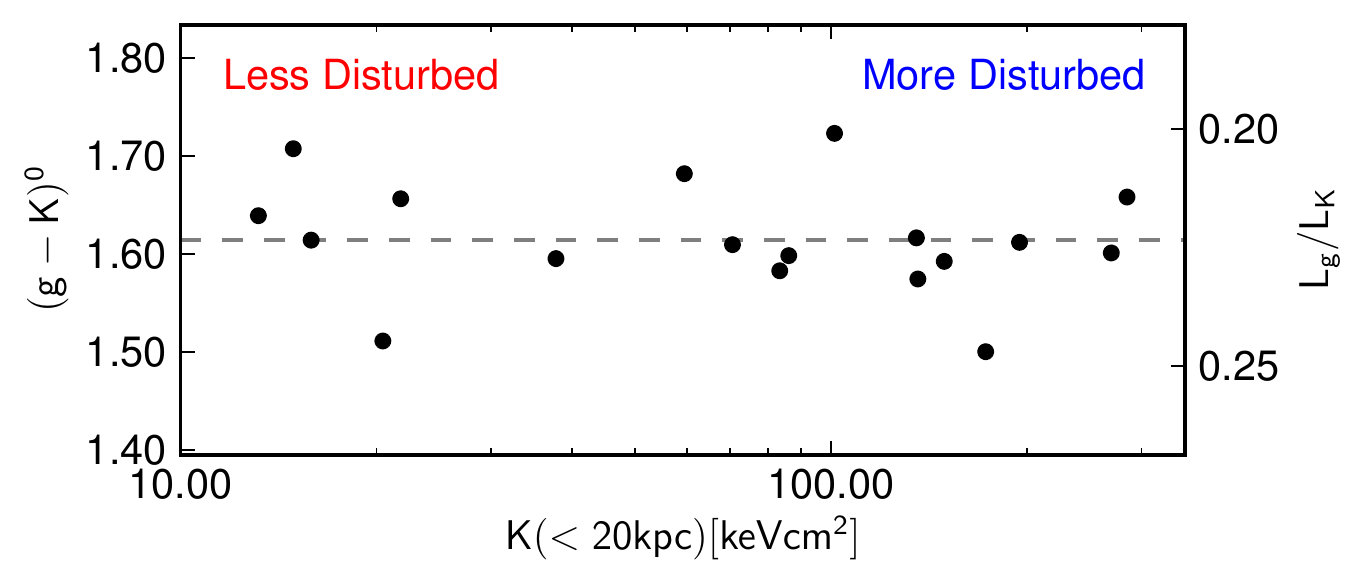}
  \caption{Rest frame ($g-K$) cluster colour within $r_{\rm vir}$ as a function of various indicators of the level of disturbance in the cluster. From top to bottom: DS statistic; magnitude gap; BCG / X-ray centroid separation; centroid shift; surface brightness concentration; alpha, the logarithmic slope of the gas density; and central entropy.}
  \label{fig:morph}
\end{figure}

In Figure \ref{fig:morph} we show the total rest frame ($g-K$) cluster colour within $r_{\rm vir}$ as a function of the seven indicators of disturbance discussed above. The bulk cluster properties ($\Delta_{DS}$, $\Delta M_{1,2}$, $\Delta^{\rm{BCG}}_{\rm{X\mbox{-}ray}}$ and $\langle w \rangle$) all show a trend of decreasing scatter ($\sigma_{(g-K)}$) in cluster colour as clusters become more disturbed. $c_{SB}$ also suggests this trend, however it does not appear in the other central cluster properties ($\alpha$ and $K$). Note that we show the values for ($g-K$) colour within $r_{\rm vir}$, because it covers the widest wavelength baseline and the whole cluster, but the results are similar with other choices of colour and radii. As an example, we show results for other colours and radii as a function of the DS statistic in the appendix (Figure \ref{fig:ds_all}).

\begin{figure}
  \centering
  \includegraphics[width=0.9\linewidth]{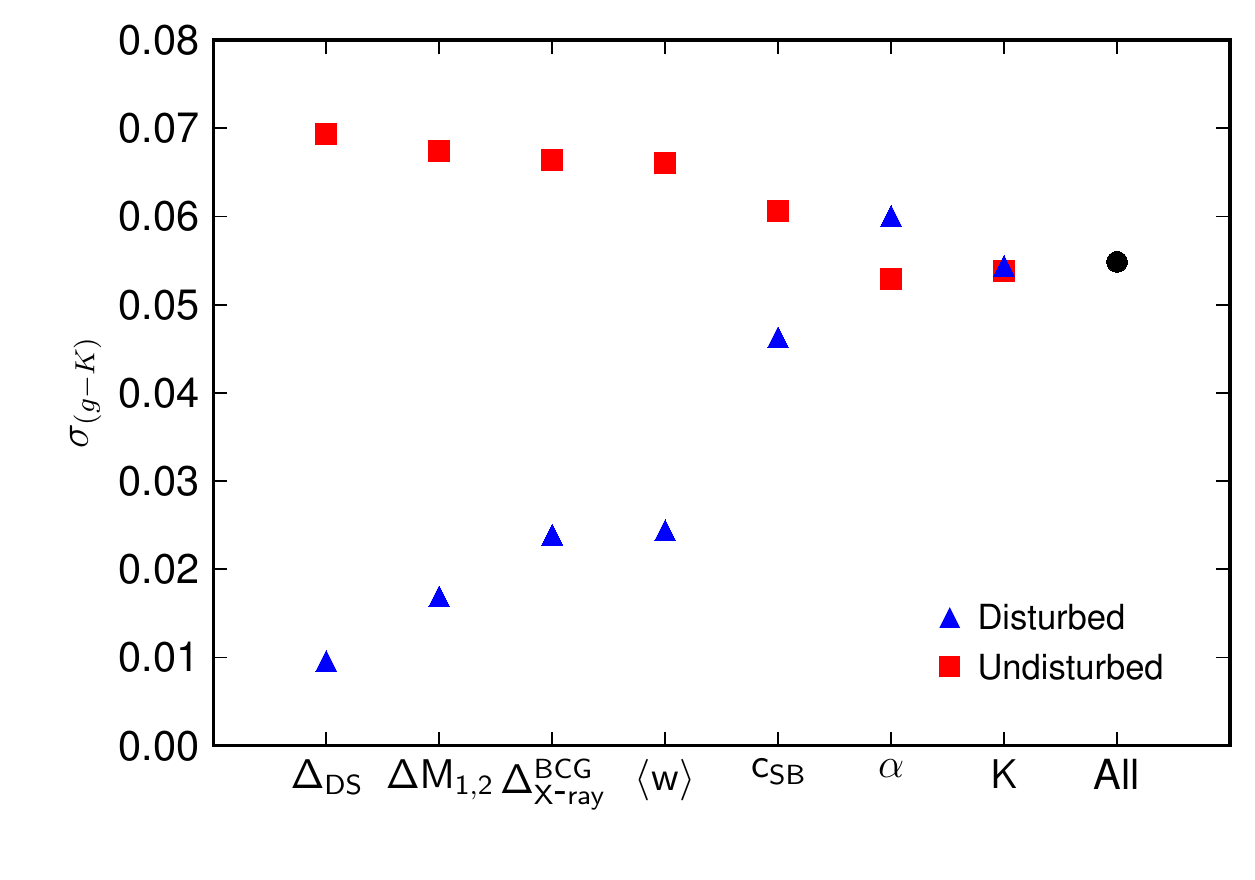}
  \caption{Variation in cluster colour within $r_{\rm vir}$, $\sigma_{(g-K)}$, for two subsamples (blue triangles: disturbed, red squares: undisturbed) defined by various indicators of the level of disturbance in the cluster.}
  \label{fig:subsamples}
\end{figure}

To quantify this trend, we split the full sample into two subsamples (disturbed and undisturbed) based on each indicator, either splitting the sample in half or where there appears to be a natural division near the median. We then calculate the spread in the cluster colour within these subsamples. As shown in Table \ref{tab:subsamples} and Figure \ref{fig:subsamples} there is a clear difference between the two subsamples for most indicators, with the disturbed clusters showing less variability in cluster colour than the undisturbed clusters. Most interestingly, the degree of variation in the subsamples varies systematically with the cluster disturbance indicator by which the sample was split. The properties towards the left of Figure \ref{fig:subsamples} are the bulk properties, thereby indicating disturbance on large scales, while those towards the right are the ICM properties and as such probe closer to the cluster centre. For instance, the farthest left parameter ($\Delta_{DS}$) measures disturbances on the scale of the whole cluster, and is often used to detect infalling galaxy groups. Similarly, the $\Delta M_{1,2}$ parameter measures disturbances within $0.5r_{\rm vir}$ (the region for which a second-rank galaxy is searched). The disturbance indicators probe smaller and smaller scales, until the right-most indicator on the figure, which probes the cluster central entropy within the central 20kpc. Taken together, these results suggest that there is a larger spread in the stellar age, metallicity and/or SFR in undisturbed clusters than in disturbed clusters, and that this effect decreases with disturbance indicators towards the cluster centre.

\begin{table}
  \caption{Variation within subsamples}\label{tab:subsamples}
  \begin{center}
    \begin{tabular}{ l c c c }
    \hline
    \hline
    Indicator & Disturbed & Undisturbed & Indicator \\
     & Clusters & Clusters & Threshold \\
    \hline
    \multicolumn{4}{c}{$\langle g - K \rangle \pm \sigma_{(g-K)}$} \\
All & \multicolumn{2}{c}{$ 1.614 \pm 0.055 $} \\
$\Delta_{DS}$ & $ 1.599 \pm 0.010 $ & $ 1.639 \pm 0.069 $ & 1.8 \\
$\sf{\Delta M_{1,2}}$ & $ 1.607 \pm 0.017 $ & $ 1.619 \pm 0.067 $ & 0.6\\
$\sf{\Delta^{BCG}_{X\mbox{-}ray}}$ & $ 1.609 \pm 0.024 $ & $ 1.617 \pm 0.066 $ & 2.5\\
$\sf{\langle w \rangle}$ & $ 1.607 \pm 0.024 $ & $ 1.618 \pm 0.066 $ & 0.6\\
$\sf{c_{SB}}$ & $ 1.603 \pm 0.046 $ & $ 1.627 \pm 0.061 $ & 0.5\\
$\sf{\alpha}$ & $ 1.610 \pm 0.060 $ & $ 1.618 \pm 0.053 $ & -0.6\\
$\sf{K}$ & $ 1.606 \pm 0.052 $ & $ 1.625 \pm 0.057 $ & 75.0\\
    \hline
    \multicolumn{4}{c}{$\langle L_g/L_K \rangle \pm \sigma_{(L_g/L_K)}$} \\
All & \multicolumn{2}{c}{$ 0.222 \pm 0.011 $} \\
$\Delta_{DS}$ & $ 0.225 \pm 0.002 $ & $ 0.217 \pm 0.014 $ & 1.8\\
$\sf{\Delta M_{1,2}}$ & $ 0.224 \pm 0.003 $ & $ 0.222 \pm 0.014 $ & 0.6\\
$\sf{\Delta^{BCG}_{X\mbox{-}ray}}$ & $ 0.223 \pm 0.005 $ & $ 0.222 \pm 0.014$ & 2.5\\
$\sf{\langle w \rangle}$ & $ 0.224 \pm 0.005 $ & $ 0.222 \pm 0.014 $ & 0.6\\
$\sf{c_{SB}}$ & $ 0.224 \pm 0.010 $ & $ 0.220 \pm 0.012 $ & 0.5\\
$\sf{\alpha}$ & $ 0.223 \pm 0.012 $ & $ 0.221 \pm 0.011 $ & -0.6\\
$\sf{K}$  & $ 0.224 \pm 0.011 $ & $ 0.220 \pm 0.012 $ & 75.0\\
    \hline
    \end{tabular}
  \end{center}
{\footnotesize}
\end{table}

\section{Interpretation}\label{sec:int}

There are three broad potential causes of variation in cluster colour. We will discuss each of these in turn:

\begin{enumerate}

\item The role of the BCG - the state of the cluster (disturbed or undisturbed) can strongly affect the colour of the BCG. It is known that undisturbed cool core clusters have BCGs with a greater range of SFRs and optical emission lines \citep{Cavagnolo2008, McDonald2010}. If the BCG colour dominated the total cluster colour, then this would lead to an increase in the variability of the cluster colour in undisturbed clusters with strongly cooling cores. While the observed trend is in the same direction, we would expect to see the biggest trend in indicators which probe near the cluster core ($\alpha$, $K$), but we see no such trend in these indicators. Additionally, the BCG luminosity is typically only $\sim 5$ per cent of the total cluster luminosity, so is subdominant.

\item Infalling galaxies - it is well known that galaxies within massive halos, such as galaxy groups and clusters, have systematically less star formation than isolated galaxies \citep[e.g.][]{McGee2011,Wetzel2012}. Furthermore, the fraction of star forming galaxies is remarkably similar in different groups \citep{Balogh2010}. Therefore, it could be the case that while undisturbed clusters are continually accreting star forming field galaxies which are quenched as they fall into the cluster, disturbed clusters are gaining their mass from infalling groups and clusters. The galaxies in these groups already reside in a dense environment, and so have already had their star formation quenched. As a result these galaxies have little impact on the overall star formation of the cluster, in contrast to the field galaxies falling into the undisturbed clusters and introducing cluster to cluster variation. This effect would decrease towards the cluster centre, consistent with the trend being clear in the bulk cluster properties but only in one of the three centre cluster properties.

We have tested this hypothesis using accretion histories for clusters from the Millennium N-body dark matter simulation \citep{Springel2005, McGee2009}. However, we find that clusters which are currently undergoing a major merger have not accreted a significantly higher fraction of their galaxies through massive halos in the last 1 - 4 Gyrs. For instance, the fraction of galaxies accreted through haloes of mass $> 10^{13}$ M$_\odot$ in the last 2 Gyrs in clusters undergoing a major merger is 0.35 $\pm$ 0.01, while it is 0.34 $\pm$ 0.01 in clusters not undergoing a major merger.

\item The effect of mergers - there has been recent evidence that major mergers may affect the star formation properties of the galaxies within clusters \citep{Rawle2014,Pranger2014,Stroe2015}. If a merger could `standardise' the SFR in a cluster, then disturbed clusters would have less variation in their total colour. As time passes since the last major merger, the spread in SFR and therefore cluster colour would increase.

The merger could standardise the SFR by leading to a burst and/or quenching, as long as it led to a similar effect in all merging clusters. One possible scenario is galaxy interaction with the shocks created by merging clusters. The Mach number of a galaxy is typically $\mathcal{M} \sim 1$ \citep{Sarazin1988}, while that of a cluster shock can be as high as 5 \citep[e.g.][]{vanWeeren2010}, so we would expect to see the effects of a shock across the entire cluster well before the galaxies were virialised. The standardisation would be seen more clearly in the cluster disturbance probes which examine the widest range (e.g. $\Delta_{DS}$), in good agreement with the observed trends we see. These cluster-wide shocks are unlikely to alter the densest gas in the cluster core, and so would not be detectable in the cluster disturbance indicators which probe the central ICM properties (e.g. $K$) \citep{Poole2008}. Any disruption to the cool core would occur during the later stages of a merger.

\end{enumerate}

Given all this, it seems that the merger itself, and perhaps the shock it triggers, is the most likely cause of the lack of variation we see in the total cluster colour of disturbed clusters. While the precise physical mechanism which causes this is unclear, upcoming low frequency radio facilities \citep[eg., LOFAR;][]{vanHaarlem2013} will find hundreds of merging clusters whose shock waves can be mapped by their radio emission, and should lead to tighter constraints on the physical mechanism.

\section{Conclusions and Implications for Future Surveys}\label{sec:summary}

In this study, we have used measurements of the luminosities and colours of 19 galaxy clusters with well measured weak-lensing masses, highly complete stellar mass limited spectroscopy, and a wide range of indicators of the levels of disturbance in the clusters. We can summarise our main conclusions as:

\begin{enumerate}
\item The slope and scatter of the relation between total cluster luminosity and cluster mass is consistent across the full range of bandpasses we probed (\textit{grizJK}). The trend in intercept of these relations is well understood if the galaxy clusters are made up of predominantly old, passive galaxies with metallicities $\sim 0.4$ solar.
\item The intrinsic scatter in these relations is $\sim 0.1$ within $r_{500}$ and suggests they would be good, cheap mass proxies for large scale photometric surveys of galaxy clusters, as discussed further below.
\item The variation in cluster colour shows trends with the overall cluster disturbance, increasing as clusters become more relaxed, perhaps indicating that the major mergers are a standardising force in the global colours, possibly through system-wide shocks.
\end{enumerate}

We have shown that total cluster luminosity scales tightly with weak-lensing mass over the full range of wavelengths considered here (Table \ref{tab:fits}, Figures \ref{fig:relations} \& \ref{fig:wavelengths}). Combined with the fact that these measurements were made on shallow survey data, this work suggests that these luminosities are promising mass proxies for future surveys, consistent with previous studies \citep[e.g.][]{Girardi2000,Lin2003,Lin2004,Ramella2004,Popesso2005,Mulroy2014,Pearson2015,Ziparo2016}. We highlight that the specific luminosity measurements used in this work benefited from highly complete $K$ band limited (roughly stellar mass limited) spectroscopic membership catalogues, and prior radial knowledge from weak-lensing analysis. Future studies will be needed to quantify the best method for luminosity measurements in the absence of this prior information.

\begin{figure}
  \centering
  \includegraphics[width=\linewidth]{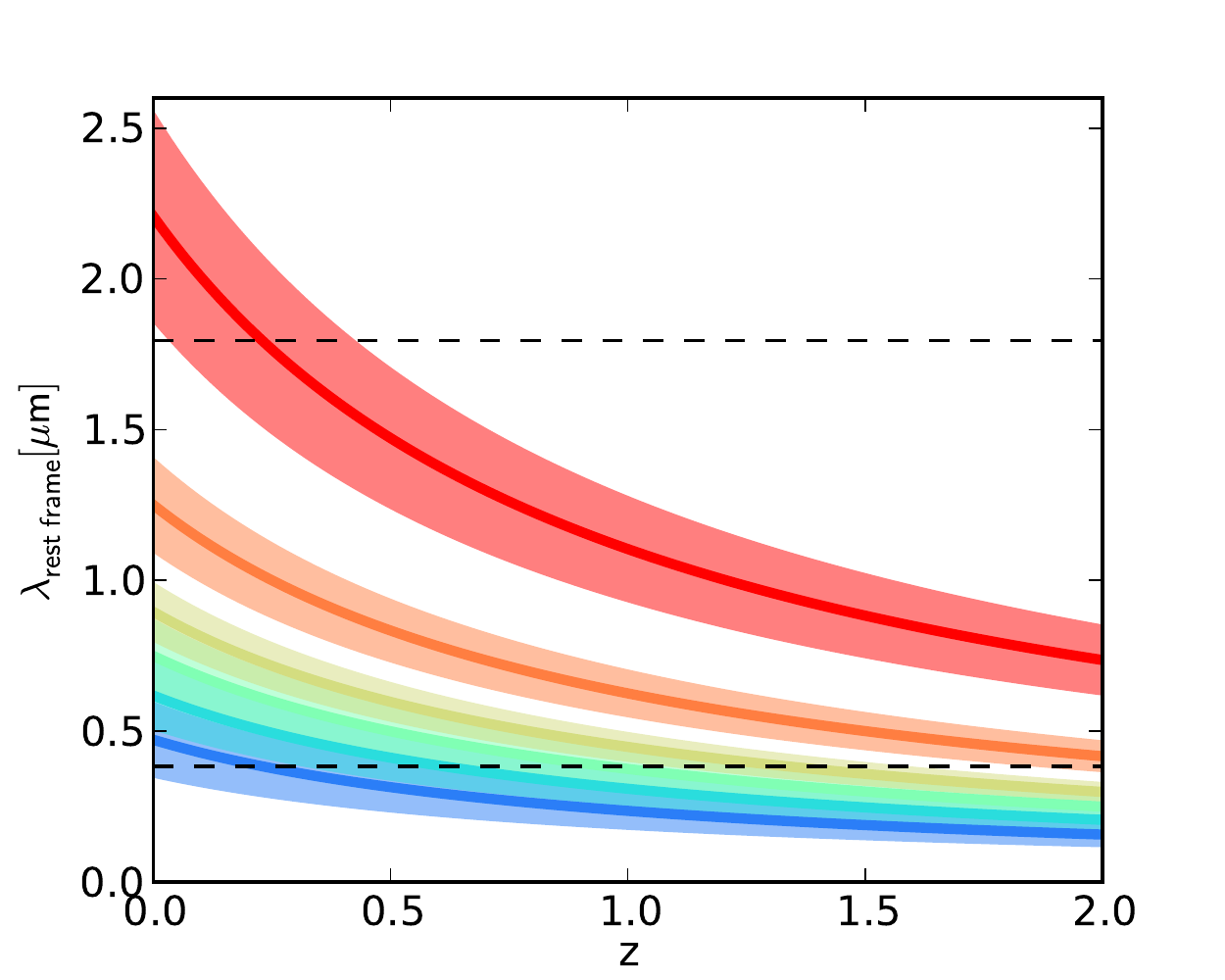}
  \caption{Rest frame wavelength for each bandpass (\textit{grizJK}, shown as blue through red) as a function of redshift.}
  \label{fig:redev}
\end{figure}

The observed wavelength range considered ($\sim 0.47 - 2.21 \mu \rm{m}$) corresponds to a rest frame wavelength range of $\sim 0.38 - 1.80 \mu \rm{m}$ at our average redshift $\langle z \rangle=0.23$, for which we have shown these luminosities to tightly scale with mass. The redshift evolution of this rest frame wavelength can be seen in Figure \ref{fig:redev}, which highlights the importance of observed near-infrared wavelengths when studying clusters at redshifts of 1 and above, which is significant for ongoing and upcoming surveys and instruments such as DES, HSC, Euclid, and LSST.

\section*{Acknowledgements}

SLM acknowledges support from an STFC Postgraduate Studentship.

\appendix

\section{Range of colours and radii}\label{sec:dev_all}

\begin{figure*}
  \centering
  \includegraphics[width=0.3\linewidth]{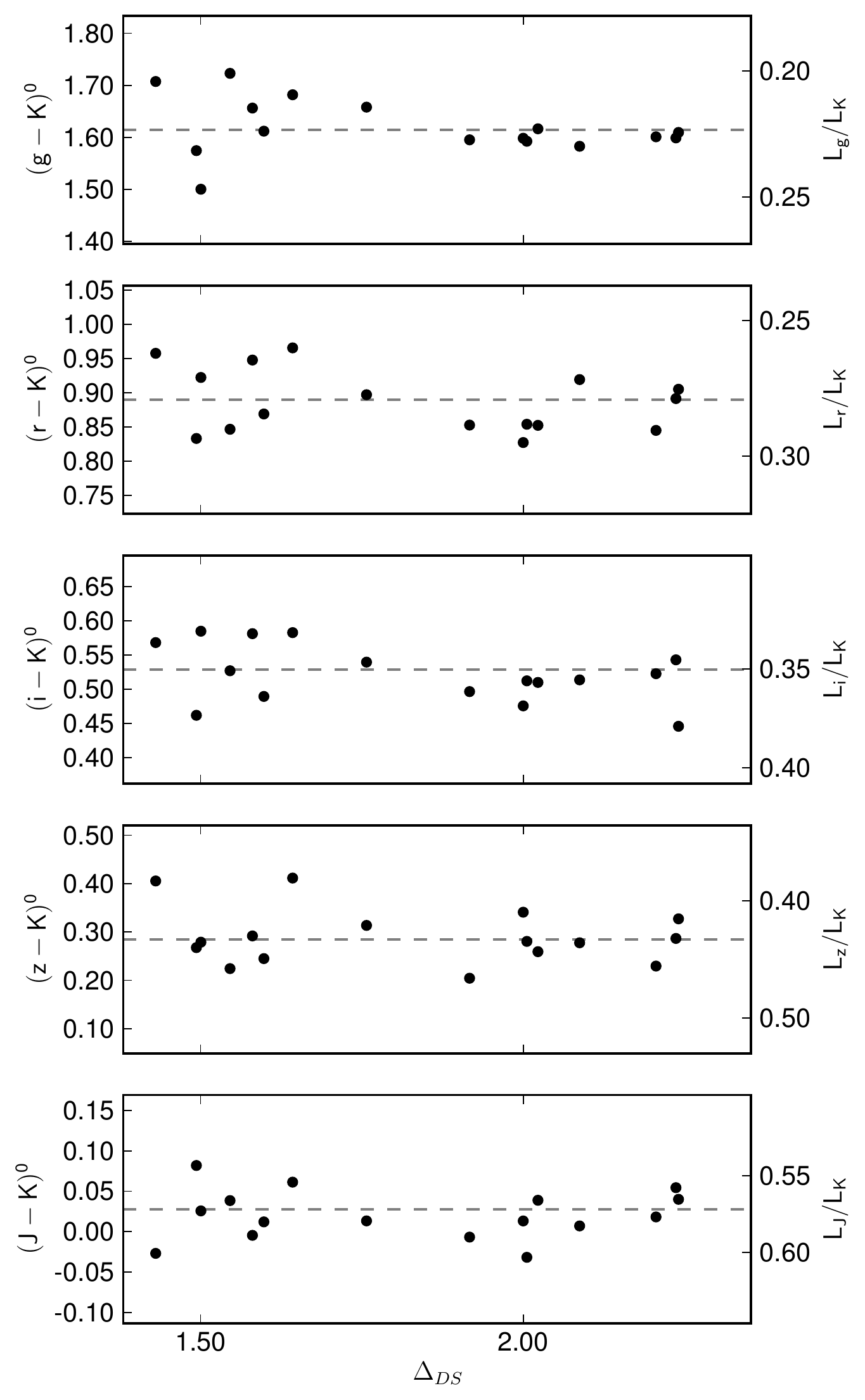}
  \includegraphics[width=0.3\linewidth]{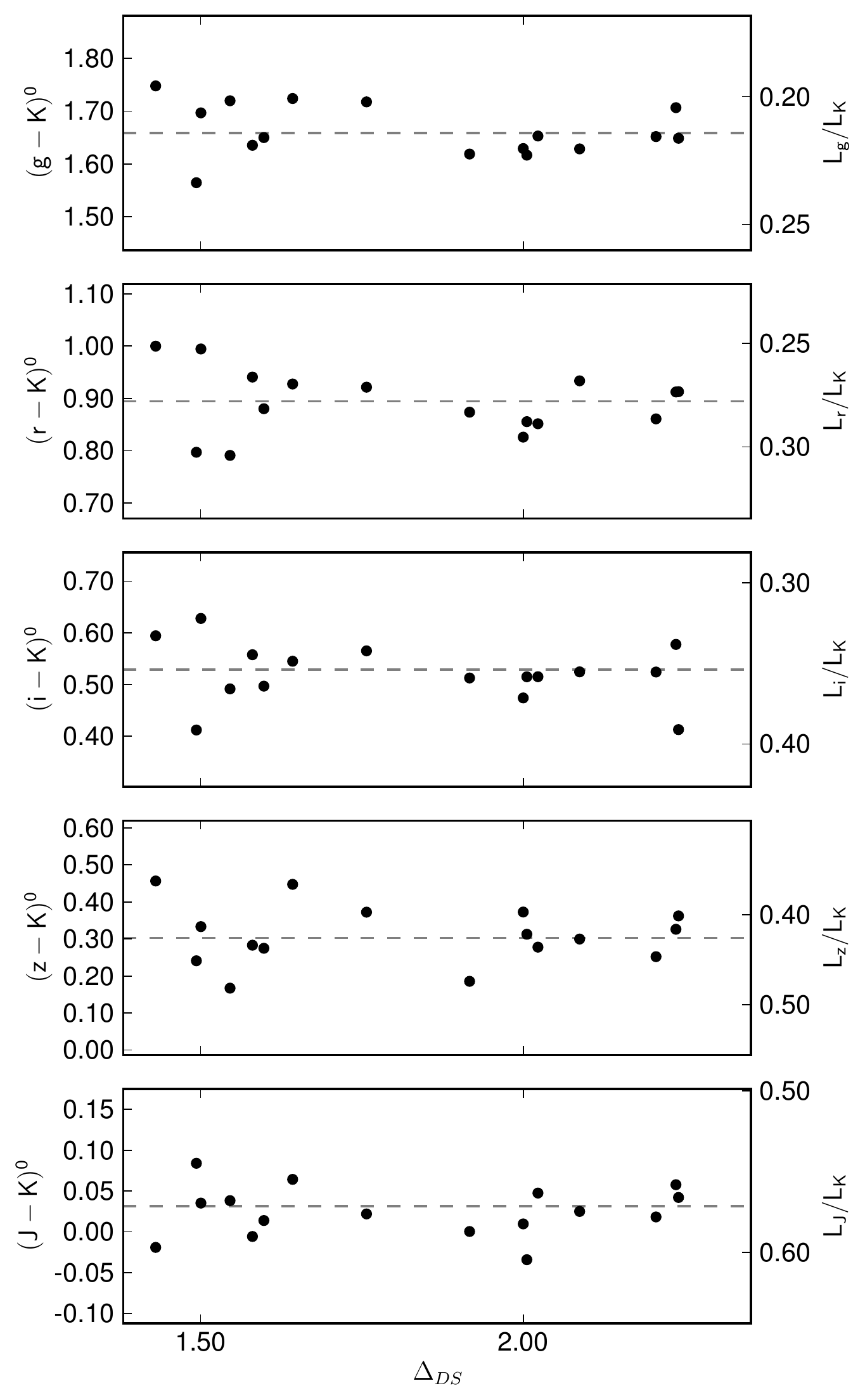}
  \includegraphics[width=0.3\linewidth]{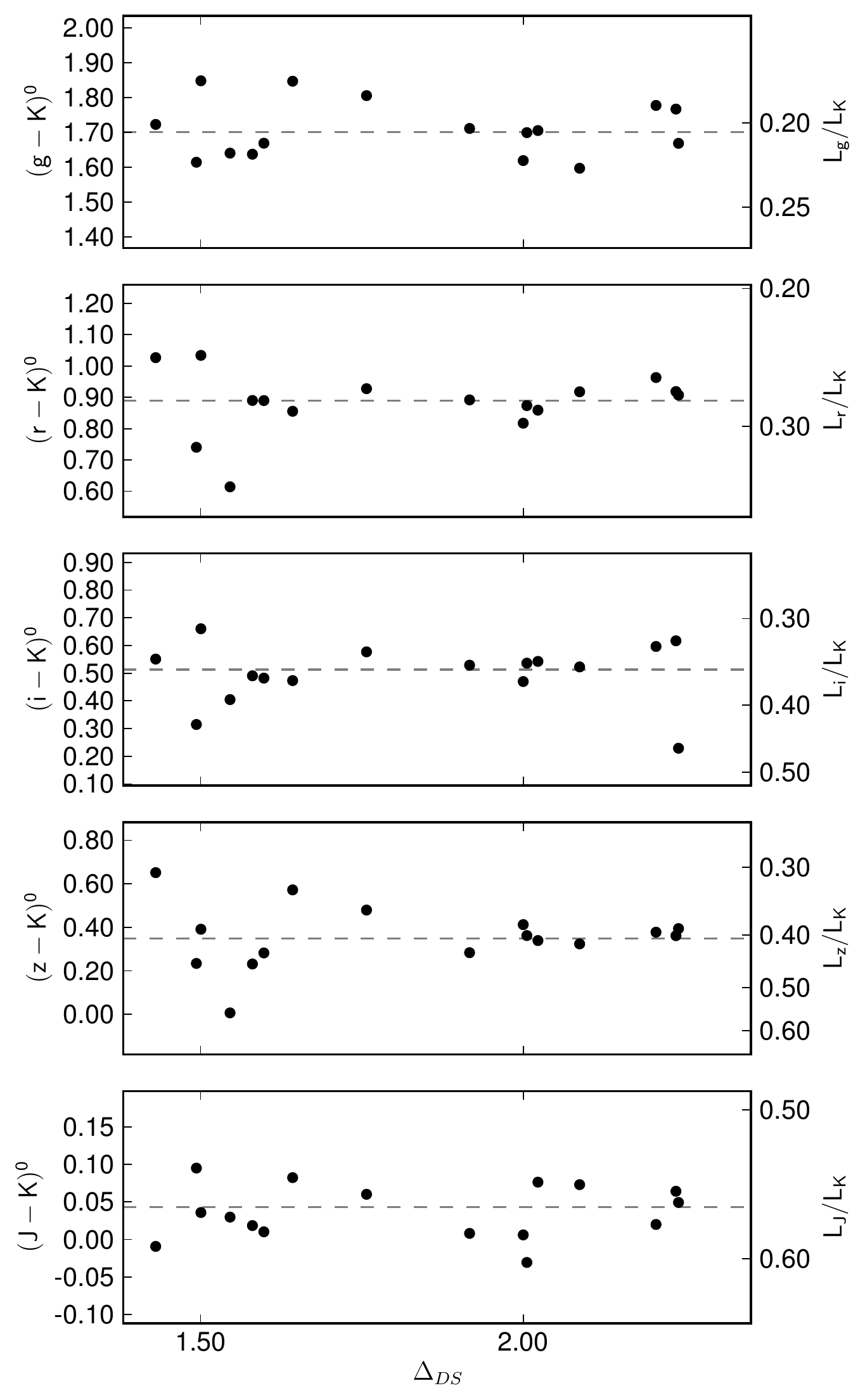}
  \caption{Rest frame ($g-K$) cluster colour within $r_{\rm vir}$ [left], $r_{500}$ [middle] and $r_{2500}$ [right] as a function of the DS statistic.}
  \label{fig:ds_all}
\end{figure*}

We show as an example in Figure \ref{fig:ds_all} the trend of cluster colour with the DS statistic, where the colour is defined over a range of colours ($g-K$, $r-K$, $i-K$, $z-K$, $J-K$) and calculated within a range of radii ($r_{\rm vir}$, $r_{500}$, $r_{2500}$). The trend seen in the top left panel and discussed in Section \ref{sec:colour_trends} is also visible in most other colour/radius combinations.

\bibliographystyle{mnras}
\bibliography{mulroybib}

\label{lastpage}
\end{document}